\documentclass[reprint, superscriptaddress, nofootinbib,aps,prl]{revtex4-1}

\pdfoutput=1
\usepackage{mathtools, amsfonts, amsthm, latexsym, amssymb,dsfont}
\usepackage[T1]{fontenc}
\usepackage{microtype}
\usepackage{times}
\usepackage{slashed}
\usepackage{hyperref}
\hypersetup{colorlinks=true, citecolor=blue, urlcolor=blue}

\newcommand{\mL}{\mathcal{L}}

\newcommand{\pd}{\partial}

\usepackage{color}

\begin{document}
\title{Phases of Wilson Lines in Conformal Field Theories}
\author{Ofer Aharony}
\email{ofer.aharony@weizmann.ac.il}      
\affiliation{Department of Particle Physics and Astrophysics, Weizmann Institute of Science, Rehovot 7610001, Israel}                                                              
\author{Gabriel Cuomo}
\email{gcuomo@scgp.stonybrook.edu}
\affiliation{Simons Center for Geometry and Physics, SUNY, Stony Brook, NY 11794, USA}
\affiliation{C. N. Yang Institute for Theoretical Physics, Stony Brook University, Stony Brook, NY 11794, USA}
\author{Zohar Komargodski}
\email{zkomargodski@scgp.stonybrook.edu}
\affiliation{Simons Center for Geometry and Physics, SUNY, Stony Brook, NY 11794, USA}
\author{M\'ark Mezei}
\email{mezei@maths.ox.ac.uk}   
 \affiliation{Mathematical Institute, University of Oxford, Woodstock Road, Oxford, OX2 6GG, United Kingdom}  
 \affiliation{Simons Center for Geometry and Physics, SUNY, Stony Brook, NY 11794, USA}
\author{Avia Raviv-Moshe}
\email{araviv-moshe@scgp.stonybrook.edu}
\affiliation{Simons Center for Geometry and Physics, SUNY, Stony Brook, NY 11794, USA}
\date{\today}

\begin{abstract}
We study the low-energy limit of Wilson lines (charged impurities) in conformal gauge theories in 2+1 and 3+1 dimensions. As a function of the representation of the Wilson line, certain defect operators can become marginal, leading to interesting renormalization group flows and for sufficiently large representations to complete or partial screening by charged fields.
This result is universal: in large enough representations, Wilson lines are screened by the charged matter fields.  We observe that the onset of the screening instability is associated with fixed-point mergers.  We study this phenomenon in a variety of applications. 
In some cases, the screening of the Wilson lines takes place by dimensional transmutation and the generation of an exponentially large scale.  
We identify the space of infrared conformal Wilson lines in weakly coupled gauge theories in 3+1 dimensions and determine the screening cloud due to bosons or fermions. We also study QED in 2+1 dimensions in the large $N_f$ limit and identify the nontrivial  conformal Wilson lines. 
We briefly discuss 't~Hooft lines in 3+1-dimensional gauge theories and find that they are screened in weakly coupled gauge theories with simply connected gauge groups. 
In non-Abelian gauge theories with S-duality,
this together with our analysis of the Wilson lines gives a compelling picture for the screening of the line operators as a function of the coupling.

\end{abstract}
\maketitle

\paragraph*{Introduction} 
A natural question in Quantum Field Theory is to understand the space of extended operators. Local operators and their correlation functions have been studied intensely, but relatively little is understood about the space of extended operators.

Relativistic invariance allows us to think of extended operators as either nonlocal operators acting at a given time, or as a modification of the Hamiltonian by an insertion of an impurity in some region of space.   Among the possible extended operators,  line (one-dimensional) defects provide the simplest examples, as they correspond to a localized impurity in the Hamiltonian picture.

Here we will consider the simplified (yet of great physical importance) scenario when the bulk theory (away from the defect) is a Conformal Field Theory (CFT). Examples of interesting line defects in CFTs include symmetry defects \cite{Calabrese:2004eu, Billo:2013jda,Gaiotto:2013nva, Giombi:2021uae}, spin impurities \cite{tsvelick1985exact,Affleck:1995ge,sachdev1999quantum,vojta2000quantum,Liu:2021nck,Cuomo:2022xgw,Beccaria:2022bcr,Nahum:2022fqw},  localized external fields \cite{Assaad:2013xua,Parisen_Toldin_2017,Cuomo:2021kfm, Popov:2022nfq,Giombi:2022vnz}, and 't Hooft and Wilson lines in conformal gauge theories.

While the bulk theory is at a fixed point of the renormalization group (RG) flow, in general, a nontrivial RG flow  takes place on the defect. One expects on general grounds an infrared fixed point of the line defect, preserving (for a straight or circular defect) the 1-dimensional conformal algebra $sl(2,\mathbb{R})$. The infrared fixed point may or may not be trivial. Defect operators are classified according to their $sl(2,\mathbb{R})$ charges. Such a system is commonly referred to as a {\it Defect Conformal Field Theory} (DCFT). See~\cite{Billo:2016cpy} for an introduction to the subject. RG flows on a defect can be triggered by perturbing a DCFT with relevant defect operators. A central question about the dynamics of line defects concerns with their infrared limit, and in particular, if the infrared is screened (i.e. furnishes a trivial DCFT) or not. 
Under the assumptions of locality and unitarity, RG flows on line defects are constrained by a monotonically decreasing entropy function \cite{Cuomo:2021rkm} (in the case of 1+1 dimensional bulk, see also \cite{Affleck:1991tk,Shatashvili:1993kk,Shatashvili:1993ps,Friedan:2003yc,Casini:2016fgb}). Another general constraint on RG flows on line defects is due to one-form symmetry: If the line operator is charged under an endable one-form symmetry then it cannot flow to a trivial defect in the infrared.\footnote{For the definitions and a review, see~\cite{Rudelius:2020orz}.} Finally, there are constraints on conformal defects due to the bootstrap equations, see~\cite{Lauria:2020emq,Collier:2021ngi,Herzog:2022jqv,Gimenez-Grau:2022czc} for recent examples and references.

In this letter,  we focus on a particular class of line operators that naturally exist in conformal gauge theories, i.e. Wilson lines  \cite{Wilson:1974sk}. A Wilson line physically describes the insertion of an (infinitely heavy) charged test particle that moves on a worldline $\gamma$:
\begin{equation}\label{eq_WilsonLineDef}
W_R (\gamma) = \mathrm{tr}_R\left(P\exp{\left(i\int_\gamma A_\mu dx^\mu\right)}\right),
\end{equation}
where $R$ is the representation of the gauge group. For a timelike $\gamma$, we can think of the Wilson line as a localized charged impurity -- it changes the Hamiltonian and the ground-state of the system. The definition \eqref{eq_WilsonLineDef} brings up the following natural question: What is the infrared limit of the Wilson line operator as a function of its representation? In particular, an intriguing property of \eqref{eq_WilsonLineDef} is that there is no continuous free parameter in the definition of the Wilson line operator.  However, as we will argue in this letter, this does not mean that no RG flow takes place.  
Physically, this is expected because the electric field sourced by a Wilson line in a sufficiently large representation may destabilize the vacuum. The primary goal of this work is to study this general phenomenon from the viewpoint of defect RG.

We discuss an instability of Wilson lines to screening by charged fields (fermions or bosons) in a variety of physical setups, and relate this to the flow of bilinear operators integrated along the line defect. These bilinear operators are in some sense ``missing'' from the definition of the Wilson line~\eqref{eq_WilsonLineDef}. We calculate the $\beta$-functions associated with the flow of these new couplings. 
For small enough charges,  the new couplings admit several interesting novel fixed points, while for supercritical charges they flow to large values and dimensional transmutation takes place on the defect. 
We observe fixed-point mergers taking place at a  critical charge. This behavior is reminiscent of how conformality is lost in some QCD-like theories \cite{Kaplan:2009kr,Gorbenko:2018ncu,Benini:2019dfy}. The dimensional transmutation associated with the fixed-point merger implies that the screening cloud is exponentially large. 

A brief, qualitative summary of our findings is as follows: We find that, at weak coupling in  3+1 dimensions, Wilson lines are nontrivial in the infrared for charges (weights) $\lesssim {1\over g_\text{YM}^2}$, while if the charge (weight) exceeds $\sim {1\over g_\text{YM}^2}$ the bilinear operators flow to strong coupling and screen the infrared partially or fully. In 2+1 dimensions the situation is qualitatively different for weakly coupled charged scalars  -- a scalar bilinear operator leads to a trivial infrared limit of the Wilson line already for small charges (weights). On the other hand a weakly coupled charged  fermion in 2+1 dimensions does not lead to immediate screening of all Wilson lines. 

These instabilities of Wilson lines are general. They were previously discussed for heavy nuclei in QED \cite{pomeranchuk1945energy} and for impurities in graphene \cite{Pereira_2007,shytov2007vacuum}. We will discuss these two cases below in more detail.

We cover several examples in this letter:
\begin{itemize}
\item Scalar (Fermion) QED$_4$: A relevant bilinear operator must be added to~\eqref{eq_WilsonLineDef} for Wilson lines with charge $|q|>{2\pi \over e^2}$ (respectively, $|q|>{4\pi \over e^2}$). The coefficient of the bilinear becomes large, and the infrared is qualitatively different from a Coulomb field with $q$ units of charge. It is completely (partially) screened by a condensate cloud, which in some cases is exponentially large.
For $|q|<{2\pi \over e^2}$ (respectively, $|q|<{4\pi \over e^2}$) the bilinear operator is irrelevant but still important to consider, since there are in general multiple UV fixed points, which lead to new DCFTs with relevant operators. 
 
\item Non-Abelian gauge theories in 3+1 dimensions: As in the QED$_4$ examples, for large enough representations, the infrared limit of the Wilson lines is either completely or partially screened. For sub-critical representations, there are potentially several fixed points corresponding to the Wilson line. 

\item QED$_3$ with $2N_f$ Dirac fermions (this is also an important example of a deconfined critical point in condensed matter physics): We find that Wilson lines up to charge $\lesssim 0.56N_f$ are not screened, while they are screened otherwise. This holds at leading order in the $1/N_f$ expansion. 
\end{itemize}

Finally, we also study similar instabilities for 't Hooft lines in four-dimensional conformal gauge theories. Combined with our analysis of the Wilson lines and with non-Abelian electric-magnetic duality, we find a compelling picture for the screening of the line operators as a function of the coupling in ${\mathcal{N}}=4$ supersymmetric Yang-Mills (SYM) and in the $SU(2)$ Seiberg-Witten ${\mathcal{N}}=2$ theory with four fundamental hypers.

An expanded version of this letter can be found in \cite{Aharony:ToAppear}, where the calculations are presented in detail and a few additional examples in 2+1 dimensions are studied.

\paragraph*{Scalar QED$_4$}

We consider massless scalar $\text{QED}$ in 3+1 dimensions with a charge $q$ Wilson line that extends in the time direction at some fixed spatial location $\vec{x}=0$. The action (in mostly minus Minkowski signature) is given by:
\begin{equation} \label{eq_Action_SQED4}
S = \int d^4x \left[-\frac{1}{4e^2}F_{\mu\nu}^2 +\left| D_\mu\phi \right|^2-\frac{\lambda}{2}|\phi|^4\right]-q\int dt A_0,\\
\end{equation}
where $A_\mu$ is the gauge field, $F_{\mu\nu}$ is the field strength, $\phi$ is a complex scalar field of charge one, $D_\mu =\partial_\mu -ieA_\mu$ is the covariant derivative, $e$ is the electric charge and $\lambda$ is a coupling constant. By rescaling  $\phi\to\phi/e$ one identifies the following double-scaling limit:\footnote{Similar double-scaling limits were recently considered in \cite{Cuomo:2022xgw,Beccaria:2022bcr,Rodriguez-Gomez:2022gbz} for different line defects.}
\begin{equation}\label{eq_doubleScaling1}
\begin{aligned}
&e\to 0, \quad \lambda\to 0, \quad q\to \infty,\\
& \frac{\lambda}{e^2}=\text{fixed},\quad q e^2=\text{fixed},
\end{aligned}
\end{equation}
in which the theory can be treated in the saddle point semiclassical approximation. The generated mass scale associated with QED (i.e. the Landau pole) becomes negligible in this limit and thus one can ignore any RG flow in the bulk and apply the formalism of DCFT. The semiclassical solution yields a Coulomb-potential solution for the gauge field $A_0 = \frac{e^2 q}{4\pi r}$ and a vanishing profile for the scalar $\phi=0$.

For reasons that will soon become clear, let us consider the defect operator $\phi^\dagger\phi$ on the line. 
Its scaling dimension can be found from
the propagator of $\phi$ fluctuations in the background $A_0 = \frac{e^2 q}{4\pi r}$. 
The defect scaling dimension is inferred from the falloff
$\phi\sim r^{{\hat \Delta_{\phi^\dagger \phi}\over 2}-1}$
of spherically symmetric solutions to the linearized equations of motion.
One finds:\footnote{\label{footsecon}
There also exists a second solution for $\hat{\Delta}_{\phi^\dagger \phi}$, whose significance will be explained below.}
\begin{equation}\label{eq_ScalarBilinearDim}
\hat \Delta_{\phi^\dagger \phi} =1+\sqrt{1-{e^4q^2\over 4\pi^2}}~. 
\end{equation}
The formula~\eqref{eq_ScalarBilinearDim}  is exact in the limit \eqref{eq_doubleScaling1}. Note that for $q=0$, $\hat \Delta_{\phi^\dagger \phi}=2$, which ought to be the case since the bulk and defect scaling dimensions coincide for a trivial defect. For small $e^2q$, the expression \eqref{eq_ScalarBilinearDim} agrees with standard Feynman diagrammatic calculations in perturbation theory.

\begin{figure}[t!]
\centering
\includegraphics[width=70mm]{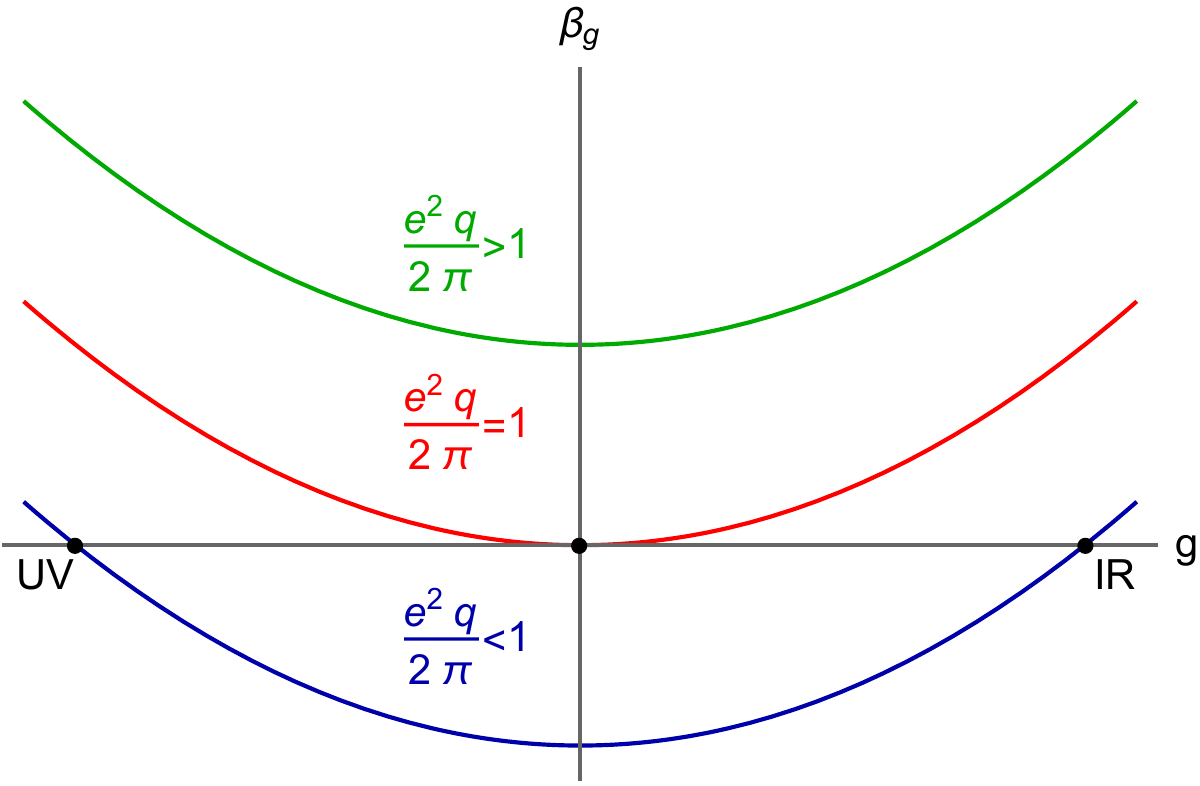}
\caption{An illustration of the $\beta$-function associated with the parameter $g$ in equation \eqref{eq_MoreGeneralLine}.  \label{fig_betaFunction}}
\end{figure}

Let us take $q>0$ without loss of generality.
The scaling dimension~\eqref{eq_ScalarBilinearDim} implies that for ${e^2q\over 2\pi}=1$ the operator becomes marginal on the defect, while for ${e^2q\over 2\pi}>1$ the scaling dimension becomes complex. Since the bilinear operator is marginal at ${e^2q\over 2\pi}=1$ and slightly irrelevant as we approach ${e^2q\over 2\pi}=1$ from below, we learn that ignoring it in RG flows is inconsistent. In other words, one must consider a more general line defect operator:
\begin{equation}\label{eq_MoreGeneralLine}
W=P \exp{\left(iq\int dtA_0-ig\int dt\, \phi^\dagger\phi\right)}.
\end{equation}
In the above, both integrals are evaluated at $\vec{x}=0$. The parameter $g$ associated with the bilinear line operator has a nontrivial beta function (that can be calculated using methods similar to those given in \cite{Aharony:2015afa}). The structure of the beta function is shown in figure \ref{fig_betaFunction}. The resulting phase diagram is analogous to the one which is found for double-trace deformations of a theory with an operator in the double-quantization window in AdS/CFT  \cite{Witten:2001ua,Klebanov:1999tb,Gubser:2002vv,Faulkner:2009wj}.

For ${e^2q\over 2\pi}<1$ there are two fixed points, corresponding to two conformal boundary conditions for the scalar near the defect. One of them yields a stable DCFT with no relevant operators, and the other gives an unstable DCFT with one relevant line operator, $\phi^\dagger\phi$. The scaling dimension of the relevant operator is given by
\begin{equation}
\hat \Delta_{\phi^\dagger \phi} =1-\sqrt{1-{e^4q^2\over 4\pi^2}}~. 
\end{equation}
Starting from the right side of the left-sided fixed point in the blue (bottom) line, one observes a double-trace-like flow from the unstable to the stable DCFT.\footnote{Consistently with this phase diagram, 
the known result for the partition function of a double-trace deformed theory \cite{Diaz:2007an,Giombi:2014xxa} implies that the $g$-function of the defect decreases under this RG flow,  in agreement with the general theorem of \cite{Cuomo:2021rkm}.} An analogous RG flow has been recently analyzed in Chern-Simons theories in the 't Hooft limit, for Wilson lines in the fundamental representation \cite{Gabai:2022vri,Gabai:2022sij}.

For ${e^2q\over 2\pi}=1$ the two fixed points merge and the defect operator $\phi^\dagger \phi$ is  marginal.  For ${e^2q\over 2\pi}>1$ the coupling $g$ flows to $-\infty$ and the infrared has to be analyzed separately. 
As mentioned in the introduction, the physical behavior described in figure \ref{fig_betaFunction} is reminiscent of how conformality is believed to be lost in QCD~\cite{Kaplan:2009kr}. Here, conformality is lost when ${e^2q\over 2\pi}=1$ in the sense that no DCFTs with finite $g$ exist for ${e^2q\over 2\pi}>1$. 

\begin{figure}[t!]
\centering
\includegraphics[width=77mm]{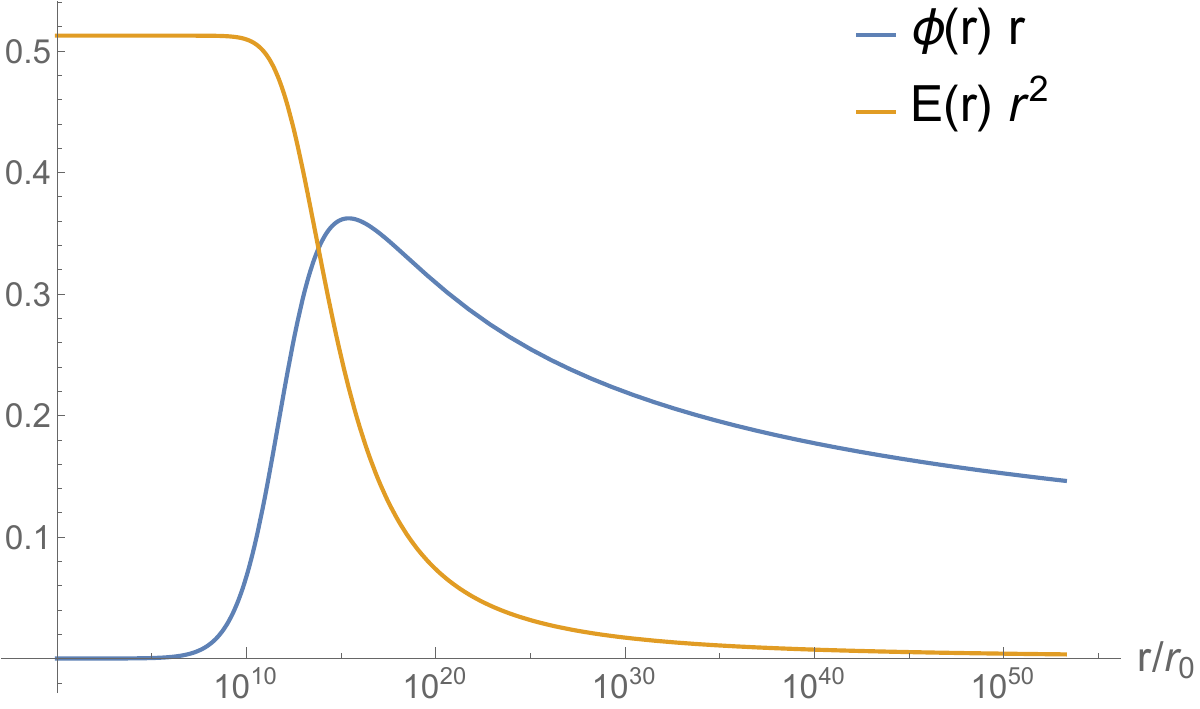}
\caption{Plots of the scalar profile (blue) and the electric field (orange) as functions of the distance from the probe charge, all normalized to be dimensionless.  The analysis was carried out for $\frac{e^2q}{2\pi}=1.02$ and $\lambda/e^2=\frac12$ by numerically solving the classical equations of motion that follow from \eqref{eq_Action_SQED4}, with boundary conditions such that the fields decay at infinity, while at a minimal radial position $r=r_0$ we have $\left.F_{0r}\right|_{r=r_0}=\frac{e^2q}{4\pi r^2}$, and $\left.|\phi|^2\right|_{r=r_0}=0$, with $\phi \neq 0$ for all $r>r_0$. Different boundary conditions for the scalar field lead to a qualitatively similar plot.
\label{fig_screening}}
\end{figure}

It is of physical interest to analyze the flow when $g\to -\infty$ in order to determine the IR behavior of Wilson lines with sufficiently large charge. Such line operators are only defined with a cutoff $r_0$, that can be viewed as the nucleus size. In this case one finds that the trivial saddle point where $\phi=0$ admits a tachyonic instability. The stable saddle point can be obtained numerically.   An example is shown in figure \ref{fig_screening}.
The electric field starts in the UV (small $r$) as a Coulomb field and decays until it is completely screened. Accordingly, the scalar profile starts at zero,  develops a cloud, and eventually gets screened as well. The integrated charge associated with the scalar condensate is exactly $-q$; i.e. the Wilson line is fully screened.
Therefore defects with ${e^2q\over 2\pi}>1$ are trivial DCFTs in the infrared.\footnote{In particular, all the bulk one-point correlation functions studied at distances much larger than the size of the cloud vanish.} The same phenomenon and screening mechanism are observed also in the case of ${e^2q\over 2\pi}<1$ if the RG flow starts in the UV from the left side of the unstable fixed point in the blue (bottom) line of figure \ref{fig_betaFunction}.

Analogously to some QCD-like theories, one finds that an exponentially low energy scale is generated when ${e^2q\over 2\pi}$ is slightly larger than $1$, and dimensional transmutation takes place.  This implies that the size of the cloud is exponentially large in units of the cutoff
\begin{equation}\label{eq_Rcloud_boson}
R_{\text{cloud}}\sim r_0 \exp\left[
\frac{2\pi}{\sqrt{\frac{e^4q^2}{4\pi^2}-1}}\right]\,.
\end{equation}
Eq. \eqref{eq_Rcloud_boson} is derived from the structure of the beta function, analogously to the correlation length in the BKT phase transition \cite{Kaplan:2009kr}.

We also comment that while we find two fixed points for ${e^2q\over 2\pi}<1$, we do not claim that our analysis of that region is complete. For ${e^2q\over 2\pi}\leq\frac{\sqrt{3}}{2}<1$  the quartic $|\phi|^4$ becomes relevant in the unstable fixed point, and the dynamics must be re-analyzed. We leave this to future research. In the range $\frac{\sqrt{3}}{2}<{e^2q\over 2\pi}<1 $ the bilinear operator in \eqref{eq_MoreGeneralLine} is the only operator that must be added and our analysis is complete in that regime.

\paragraph*{Fermion QED$_4$} 
We consider massless fermionic QED in 3+1 dimensions in the presence of a Wilson line of charge $q>0$  extending in the time direction. The action is given by:
\begin{equation} \label{eq_Action_fer_main4d}
S = \frac{1}{e^2}\int d^4x \left[-\frac{1}{4}F_{\mu\nu}^2+i\bar{\psi}_D\slashed{D}\psi_D  \right]-q\int dt A_0
\end{equation}
where $\psi_D$ is a Dirac spinor in four dimensions with charge $1$ under the $U(1)$ gauge group.
We again work in the semiclassical regime specified by the double-scaling limit in which $e \to 0$, $q\to \infty$ and $e^2 q = \text{fixed}$.

The classical saddle point is $A_0 = \frac{e^2q}{4\pi r}$, $\psi_D=0$. Expanding around it we find that there are four spin $1/2$ modes that admit two conformal boundary conditions each, for charges such that $\frac{\sqrt{3}}{2}< \frac{e^2q}{4\pi }<1$. Criticality occurs for $q=q_c={4\pi \over e^2} $ (which gives $q_c\simeq 137$ in nature \cite{pomeranchuk1945energy}).  Due to the four independent modes, there are $16$ independent bilinear operators that must be taken into account and added to the defect action. These $16$ operators can be conveniently classified according to their parity, axial and $SU(2)$ spin charges; only one bilinear operator preserves all the symmetries.

The resulting phase diagram is a generalization of that shown in figure \ref{fig_betaFunction}. In the subcritical regime, there is one unstable and one stable fixed point which preserve all the symmetries, and various unstable mixed-boundary conditions fixed points breaking some of the symmetries. These are connected by double-trace-like RG flows. Analogously to the flow from the left side of the unstable fixed point in the blue (bottom) line of figure \ref{fig_betaFunction}, there exist also runaway flows for subcritical charges. However, differently from the bosonic case, these flows only lead to screening of up to four units of charge (due to the four independent modes mentioned above).  This is a consequence of the Pauli exclusion principle, which forbids the filling of a single state with more than one fermion.
Similar to scalar QED$_4$, our analysis of the regime below criticality is not complete, e.g. for $ \frac{e^2q}{4\pi }<\frac{\sqrt{15}}{4}$ a $4$-fermion term becomes relevant near the unstable fixed points and the dynamics must be re-analyzed.

It is physically interesting to ask what is the deep IR behavior in the supercritical regime when $q>q_c$.  Unlike the scalar case, when $q>q_c$ the instability manifests itself in terms of diving states \cite{Greiner:1985ce}. Physically, these are states which change their nature from particles to holes as we raise $q$ from below to above $q_c$.  Since all hole-states must be filled in the ground-state,  the vacuum develops $q-q_c$ units of screening charge \cite{GARTNER1980181}.  For $ \sqrt{\frac{e^4q^2}{16\pi^2}-1}\ll 1$, each four units of screening charge are localized on successive shells exponentially separated from one another. One may thus account for the backreaction of the Coulomb field perturbatively and compute the radius of the exponentially large fermionic cloud \cite{shytov2007vacuum}. We find\footnote{For a massive Dirac field, we further need to demand that the radius \eqref{eq_Rcloud_fermion} is smaller than the Compton wavelength for the cloud to form. Because of that, one finds that for electrons the critical charge to screen a nucleus is $q_c\approx 175 $ \cite{pomeranchuk1945energy}. The discrepancy with the value $q_c=137$ for massless fields is thus a consequence of dimensional transmutation.}
\begin{equation}\label{eq_Rcloud_fermion}
R_{\text{cloud}}
\simeq r_0\exp\left[\frac{2\pi^2}{e^2 }\sqrt{\frac{q-q_c}{2q_c}}\right]~.
\end{equation}
As in the scalar case, the screening cloud is exponentially large due to dimensional transmutation.

\paragraph*{Non-Abelian Gauge Theories} 
Let us now discuss the implications of our findings for Wilson lines in non-Abelian conformal gauge theories in 3+1 dimensions. To be concrete, we will refer to either $SU(2)$ gauge theory with maximal $\mathcal{N}=4$ supersymmetry or the $\mathcal{N}=2$ $SU(2)$ Seiberg-Witten theory with four fundamental hypers (and for simplicity we take a vanishing $\theta$ angle). (The analysis of more general non-Abelian conformal gauge theories is analogous.) Both theories have a coupling constant $g_\text{YM}^2$ which can be chosen at will since it is an exactly marginal parameter. We consider a Wilson loop in the $(2s+1)$-dimensional representation of $SU(2)$ 
\begin{equation}\label{nAWL}
W_s={\rm Tr}\left[P e^{i \int dx^\mu A_\mu^aT^a}\right]~.
\end{equation}
This Wilson line is not BPS and it preserves the full continuous flavor symmetry of the model (unlike the BPS lines). Note that BPS lines preserve supersymmetry, which ensures that the ground state has zero energy. Hence there is no instability associated with BPS lines.\footnote{At weak coupling, non-supersymmetric Wilson lines in small representations flow to BPS lines via an $SO(6)$-breaking deformation \cite{Polchinski:2011im,Beccaria:2017rbe,Beccaria:2022bcr}.}
To analyze the phases of the line operator \eqref{nAWL} at weak coupling, we take the double scaling limit $g_\text{YM}^2\to 0$ with $g_\text{YM}^2s={\rm fixed}$ (analogously to~\eqref{eq_doubleScaling1}).
Repeating the analysis of the fluctuations around this saddle point, we find that a massless scalar field in the spin $S$ representation of the $SU(2)$ gauge group leads to complex scaling dimensions on the line defect for $g^2_\text{YM} sS> 2\pi$. Similar results hold for charged fermions and non-Abelian vector bosons. For $g^2_\text{YM} sS< 2\pi$ several fixed points, as in figure \ref{fig_betaFunction}, exist. Their characterization depends on the matter content. 

We see that at weak coupling, there is a finite but large number (of order $1/g^2_\text{YM}$) of distinct conformal line defects preserving the full flavor symmetry. As the coupling is increased, presumably only a few Wilson lines remain as nontrivial conformal line operators. 

In the $\mathcal{N}=4$ SYM theory with gauge group $SU(2)$, due to the $\mathbb{Z}_2$  one-form symmetry, the line with $s=1/2$ cannot be screened and must flow to a conformal line defect. However, in $SO(3)$ gauge theory it is possible that all Wilson lines are in fact screened at strong coupling. In the $\mathcal{N}=2$ Seiberg-Witten theory there is no one-form symmetry, and it is reasonable to expect that all the Wilson lines are screened at strong coupling.

\paragraph*{'t Hooft Lines} 

It is interesting to ask if the expectations of the previous paragraphs are compatible with S-duality in these $SU(2)$ gauge theories. 
To this end we now make some comments about (non-supersymmetric) 't Hooft lines at weak coupling. In a $U(1)$ gauge theory, a charge $q$, Lorentz spin $\ell$ particle moving around magnetic flux $n$ encounters the centrifugal barrier \cite{Shnir:2005vvi}
\begin{equation}\label{barrier}
V=\frac12|nq|{1-g\ell\over r^2}
\end{equation} 
where $g$ is the magnetic moment. The formula~\eqref{barrier} applies when $|nq|/2-\ell\geq0$. For scalars we have a repulsive centrifugal force. For fermions with the standard (weak coupling) magnetic moment $g = 2$ the numerator vanishes and we have no centrifugal barrier, leading to the familiar fermion zero modes, which correspond to marginal bilinear defect operators.

Now let us discuss charged vector bosons. In weakly coupled gauge theories they have $g = 2$. In the
case that $|nq| \geq 2$, equation~\eqref{barrier} is valid and we clearly see that the potential is attractive
with coefficient $-\frac12|nq|$, which leads to a fall-to-the-center instability of the vector bosons and conjecturally screens the 't Hooft lines. Equivalently, defect operators which are quadratic in the gauge field would have no zero for their beta function and flow to strong coupling, analogously to the behavior of scalars on the green (top) branch of figure~\ref{fig_betaFunction}.
In $\mathcal{N} = 4$ SYM theory with gauge group $SU(2)$ the minimal monopole has $|n| = 1$ and the vector boson charge is $|q|= 2$. Therefore, even the minimal ’t Hooft line is unstable to W-boson condensation at weak coupling, and deep in the infrared it presumably becomes trivial. By contrast, with gauge group $SO(3)_+$ the charge of the W-boson is 1, and hence in the background of the minimal ’t Hooft line $|n| = 1$ we have no vector boson instability, and the minimal ’t Hooft line should furnish a healthy conformal defect. (It is important that the gauge theory is $SO(3)_+$ for the minimal 't Hooft line to exist~\cite{Aharony:2013hda}.) ’t Hooft lines with $|n|>1$ are all unstable to W-boson condensation, though. These results are consistent with the magnetic one-form symmetries of the $SU(2)$ and $SO(3)_+$ theories. In summary, recalling that S-duality in ${\mathcal{N}}=4$ SYM exchanges the $SU(2)$ and $SO(3)_+$ gauge groups, the absence of 't Hooft lines at weak coupling is precisely dual to our expectations for the screening of Wilson lines as the coupling becomes strong. 

For the Seiberg-Witten ${\mathcal{N}}=2$ theory with gauge group $SU(2)$ (which is S-dual to itself), there are again no 't Hooft lines at weak coupling, and we expect no unscreened Wilson lines at strong coupling either.  For more general gauge groups, the labelling of Wilson and 't Hooft lines is explained in~\cite{Kapustin:2005py}. It would be interesting to develop an understanding of which lines are screened as a function of the coupling and the $\theta$ angle.

\paragraph*{$2+1$ Dimensional Critical Points} 
The physics of Wilson lines is of interest in 2+1 dimensions both from the particle theory point of view and also due to the existence of deconfined critical points. We will present here the physics of Wilson lines at the critical point of QED$_3$ with $2N_f$ charge 1 Dirac fermions. This fixed point is the infrared limit of the Lagrangian 
\begin{equation}\label{eq_QED3}
\mL=\frac{-1}{4e^2}F^2+i\sum_{a=1}^{2N_f}\bar{\Psi}_a\left(\slashed{\pd}-i\slashed{A}\right)\Psi_a\,.
\end{equation}
This theory has $U(1)_T\times SU(2N_f)$ global symmetry as well as time reversal symmetry. ($U(1)_T$ stands for the monopole symmetry.)
For sufficiently large $N_f$,~\eqref{eq_QED3} flows to an infrared fixed point \cite{Pisarski:1984dj}, where the gauge kinetic term is irrelevant.   In the presence of a charge $q$ Wilson line, integrating out the fermions, we have the following effective (Euclidean) action
\begin{equation}\label{Eaction}
S=-2N_f{\rm Tr}\log(\slashed{\partial}-i\slashed{A})+iq\int d \tau A_0~.
\end{equation} 
The saddle point in the presence of the Wilson line is fixed up to conformal invariance $F_{\tau i} =iE {x^i\over |x|^3}$, where $E$ is some function of $q,N_f$. Since for large $N_f$ the fermions are approximately free, we can determine when the fermions become unstable by treating them as free fields propagating in the background  $F_{\tau i} =iE {x^i\over |x|^3}$. This again requires expanding the fermions in fluctuations around the saddle point and reading out the dimension of fermion bilinears from the falloff of the fluctuations. We find that the scaling dimension of fermion bilinears on the line defect is $\Delta=1+\sqrt{1-4E^2}$ and hence the saddle point  $F_{\tau i} =iE {x^i\over |x|^3}$ is self-consistent only for $|E|\leq 1/2$. For $q\ll N_f$ we expect $E\ll1$ and hence one can solve for the saddle point by linearizing the determinant in~\eqref{Eaction}. One finds 
\begin{equation}\label{smallE} 
E= {4q\over \pi N_f} +{\mathcal{O}}\left(q^2\over N_f^2\right)~.
\end{equation} 
It is more difficult to find the answer for arbitrary $q/N_f\sim 1$. Numerically solving the saddle point equation of~\eqref{Eaction} we find that ${|q_c|\over N_f}\simeq 0.56$, i.e. Wilson lines are not screened for $|q|\leq |q_c|$ (at leading order in $1/N_f$).

The massive phases of QED$_3$ are $U(1)_{\pm N_f}$ Chern-Simons theory, which admit $N_f$ lines with nontrivial mutual braiding. It is therefore tempting to assume that the $|q|\leq |q_c|$ conformal lines at the critical point, of which there are (slightly) more than $N_f$, become the topologically nontrivial lines in the massive phases. 
Another general lesson from this example is that Wilson lines in 2+1  dimensional theories with small values of $N_f$ and $k$ would typically have few (or no) conformal Wilson lines. This is analogous to the screening of Wilson lines as the coupling is made strong in 3+1d.

The critical value $E=1/2$ is general for weakly coupled fermions in $2+1$ dimensions. In particular it also applies to fermions living on a $2+1$ dimensional plane coupled to a four-dimensional gauge field: a setup which famously describes the low energy limit of graphene \cite{kotov2012electron}. Due to the enhanced Coulomb coupling of these quasi-particles, $E=1/2$ corresponds to $q\sim 3$ \cite{Pereira_2007,shytov2007vacuum}; remarkably, this was experimentally confirmed in \cite{wang2013observing}. Our findings additionally suggest that the charge impurity admits a phase diagram analogous to the one discussed in QED$_4$, including the existence of new UV fixed points and runaway flows at subcritical $q$. 

As a final comment, we notice that for a weakly coupled charged scalar in a Coulomb field background the trivial saddle-point is always unstable in 2+1 dimensions. This is because the free bulk scaling dimension of the scalar bilinear is 1, hence the scalar sits at the fixed-point merger already at $q=0$.\footnote{Indeed, the bilinear defect perturbation of free field theory is marginally irrelevant~\cite{Cuomo:2021kfm}.}

These facts about conformal lines in bosonic and fermionic 2+1 dimensional theories could be important for 3d dualities, in the spirit of \cite{Gabai:2022vri,Gabai:2022sij}. We leave this subject for the future.

\begin{acknowledgments}
\paragraph*{Acknowledgements}
We thank S. Bolognesi, I. Klebanov, M. Metlitski, S. Sachdev, N. Seiberg, A. Sever, S. Shao, Y. Wang and S. Yankielowicz for useful discussions. We are particularly grateful to J. Maldacena for useful comments on 't Hooft lines.
The work of OA was supported in part  by an Israel Science Foundation (ISF) center for excellence grant (grant number 2289/18), by ISF grant no. 2159/22, by Simons Foundation grant 994296 (Simons
Collaboration on Confinement and QCD Strings), by grant no. 2018068 from the United States-Israel Binational Science Foundation (BSF), by the Minerva foundation with funding from the Federal German Ministry for Education and Research, by the German Research Foundation through a German-Israeli Project Cooperation (DIP) grant ``Holography and the Swampland'', and by a research grant from Martin Eisenstein. OA is the Samuel Sebba Professorial Chair of Pure and Applied Physics. 
GC is supported by the Simons Foundation (Simons Collaboration on the Non-perturbative Bootstrap) grants 488647 and 397411.  ZK, MM and ARM are supported in part by the Simons Foundation grant 488657 (Simons Collaboration on the Non-Perturbative Bootstrap) and the BSF grant no. 2018204.  ARM is an awardee of the Women’s
Postdoctoral Career Development Award. The work of ARM was also supported in part by the Zuckerman-CHE STEM Leadership Program.  
\end{acknowledgments}

\bibliography{Biblio}

\providecommand{\href}[2]{#2}\begingroup\raggedright\begin{thebibliography}{10}

\bibitem{Calabrese:2004eu}
P.~Calabrese and J.~L. Cardy, \emph{{Entanglement entropy and quantum field
  theory}}, \href{https://doi.org/10.1088/1742-5468/2004/06/P06002}{\emph{J.
  Stat. Mech.} {\bfseries 0406} (2004) P06002}
  [\href{https://arxiv.org/abs/hep-th/0405152}{{\ttfamily hep-th/0405152}}].

\bibitem{Billo:2013jda}
M.~Bill\'o, M.~Caselle, D.~Gaiotto, F.~Gliozzi, M.~Meineri and R.~Pellegrini,
  \emph{{Line defects in the 3d Ising model}},
  \href{https://doi.org/10.1007/JHEP07(2013)055}{\emph{JHEP} {\bfseries 07}
  (2013) 055} [\href{https://arxiv.org/abs/1304.4110}{{\ttfamily 1304.4110}}].

\bibitem{Gaiotto:2013nva}
D.~Gaiotto, D.~Mazac and M.~F. Paulos, \emph{{Bootstrapping the 3d Ising twist
  defect}}, \href{https://doi.org/10.1007/JHEP03(2014)100}{\emph{JHEP}
  {\bfseries 03} (2014) 100} [\href{https://arxiv.org/abs/1310.5078}{{\ttfamily
  1310.5078}}].

\bibitem{Giombi:2021uae}
S.~Giombi, E.~Helfenberger, Z.~Ji and H.~Khanchandani, \emph{{Monodromy defects
  from hyperbolic space}},
  \href{https://doi.org/10.1007/JHEP02(2022)041}{\emph{JHEP} {\bfseries 02}
  (2022) 041} [\href{https://arxiv.org/abs/2102.11815}{{\ttfamily
  2102.11815}}].

\bibitem{tsvelick1985exact}
A.~Tsvelick and P.~Wiegmann, \emph{Exact solution of the multichannel kondo
  problem, scaling, and integrability},
  \href{https://doi.org/10.1007/BF01017853}{\emph{Journal of Statistical
  Physics} {\bfseries 38} (1985) 125}.

\bibitem{Affleck:1995ge}
I.~Affleck, \emph{{Conformal field theory approach to the Kondo effect}},
  {\emph{Acta Phys. Polon. B} {\bfseries 26} (1995) 1869}
  [\href{https://arxiv.org/abs/cond-mat/9512099}{{\ttfamily
  cond-mat/9512099}}].

\bibitem{sachdev1999quantum}
S.~Sachdev, C.~Buragohain and M.~Vojta, \emph{Quantum impurity in a nearly
  critical two-dimensional antiferromagnet},
  \href{https://doi.org/10.1126/science.286.5449.2479}{\emph{Science}
  {\bfseries 286} (1999) 2479}
  [\href{https://arxiv.org/abs/cond-mat/0004156}{{\ttfamily
  cond-mat/0004156}}].

\bibitem{vojta2000quantum}
M.~Vojta, C.~Buragohain and S.~Sachdev, \emph{Quantum impurity dynamics in
  two-dimensional antiferromagnets and superconductors},
  \href{https://doi.org/10.1103/PhysRevB.61.15152}{\emph{Physical Review B}
  {\bfseries 61} (2000) 15152}
  [\href{https://arxiv.org/abs/cond-mat/9912020}{{\ttfamily
  cond-mat/9912020}}].

\bibitem{Liu:2021nck}
S.~Liu, H.~Shapourian, A.~Vishwanath and M.~A. Metlitski, \emph{{Magnetic
  impurities at quantum critical points: Large-N expansion and connections to
  symmetry-protected topological states}},
  \href{https://doi.org/10.1103/PhysRevB.104.104201}{\emph{Phys. Rev. B}
  {\bfseries 104} (2021) 104201}
  [\href{https://arxiv.org/abs/2104.15026}{{\ttfamily 2104.15026}}].

\bibitem{Cuomo:2022xgw}
G.~Cuomo, Z.~Komargodski, M.~Mezei and A.~Raviv-Moshe, \emph{{Spin impurities,
  Wilson lines and semiclassics}},
  \href{https://doi.org/10.1007/JHEP06(2022)112}{\emph{JHEP} {\bfseries 06}
  (2022) 112} [\href{https://arxiv.org/abs/2202.00040}{{\ttfamily
  2202.00040}}].

\bibitem{Beccaria:2022bcr}
M.~Beccaria, S.~Giombi and A.~A. Tseytlin, \emph{{Wilson loop in general
  representation and RG flow in 1D defect QFT}},
  \href{https://doi.org/10.1088/1751-8121/ac7018}{\emph{J. Phys. A} {\bfseries
  55} (2022) 255401} [\href{https://arxiv.org/abs/2202.00028}{{\ttfamily
  2202.00028}}].

\bibitem{Nahum:2022fqw}
A.~Nahum, \emph{{Fixed point annihilation for a spin in a fluctuating field}},
  \href{https://doi.org/10.1103/PhysRevB.106.L081109}{\emph{Phys. Rev. B}
  {\bfseries 106} (2022) L081109}
  [\href{https://arxiv.org/abs/2202.08431}{{\ttfamily 2202.08431}}].

\bibitem{Assaad:2013xua}
F.~F. Assaad and I.~F. Herbut, \emph{{Pinning the order: the nature of quantum
  criticality in the Hubbard model on honeycomb lattice}},
  \href{https://doi.org/10.1103/PhysRevX.3.031010}{\emph{Phys. Rev. X}
  {\bfseries 3} (2013) 031010}
  [\href{https://arxiv.org/abs/1304.6340}{{\ttfamily 1304.6340}}].

\bibitem{Parisen_Toldin_2017}
F.~{Parisen Toldin}, F.~F. {Assaad} and S.~{Wessel}, \emph{{Critical behavior
  in the presence of an order-parameter pinning field}},
  \href{https://doi.org/10.1103/PhysRevB.95.014401}{\emph{\prb} {\bfseries 95}
  (2017) 014401} [\href{https://arxiv.org/abs/1607.04270}{{\ttfamily
  1607.04270}}].

\bibitem{Cuomo:2021kfm}
G.~Cuomo, Z.~Komargodski and M.~Mezei, \emph{{Localized magnetic field in the
  O(N) model}}, \href{https://doi.org/10.1007/JHEP02(2022)134}{\emph{JHEP}
  {\bfseries 02} (2022) 134}
  [\href{https://arxiv.org/abs/2112.10634}{{\ttfamily 2112.10634}}].

\bibitem{Popov:2022nfq}
F.~K. Popov and Y.~Wang, \emph{{Non-perturbative defects in tensor models from
  melonic trees}}, \href{https://doi.org/10.1007/JHEP11(2022)057}{\emph{JHEP}
  {\bfseries 11} (2022) 057}
  [\href{https://arxiv.org/abs/2206.14206}{{\ttfamily 2206.14206}}].

\bibitem{Giombi:2022vnz}
S.~Giombi, E.~Helfenberger and H.~Khanchandani, \emph{{Line Defects in
  Fermionic CFTs}},  \href{https://arxiv.org/abs/2211.11073}{{\ttfamily
  2211.11073}}.

\bibitem{Billo:2016cpy}
M.~Bill\`o, V.~Gon\c{c}alves, E.~Lauria and M.~Meineri, \emph{{Defects in
  conformal field theory}},
  \href{https://doi.org/10.1007/JHEP04(2016)091}{\emph{JHEP} {\bfseries 04}
  (2016) 091} [\href{https://arxiv.org/abs/1601.02883}{{\ttfamily
  1601.02883}}].

\bibitem{Cuomo:2021rkm}
G.~Cuomo, Z.~Komargodski and A.~Raviv-Moshe, \emph{{Renormalization Group Flows
  on Line Defects}},
  \href{https://doi.org/10.1103/PhysRevLett.128.021603}{\emph{Phys. Rev. Lett.}
  {\bfseries 128} (2022) 021603}
  [\href{https://arxiv.org/abs/2108.01117}{{\ttfamily 2108.01117}}].

\bibitem{Affleck:1991tk}
I.~Affleck and A.~W.~W. Ludwig, \emph{{Universal noninteger 'ground state
  degeneracy' in critical quantum systems}},
  \href{https://doi.org/10.1103/PhysRevLett.67.161}{\emph{Phys. Rev. Lett.}
  {\bfseries 67} (1991) 161}.

\bibitem{Shatashvili:1993kk}
S.~L. Shatashvili, \emph{{Comment on the background independent open string
  theory}}, \href{https://doi.org/10.1016/0370-2693(93)90537-R}{\emph{Phys.
  Lett. B} {\bfseries 311} (1993) 83}
  [\href{https://arxiv.org/abs/hep-th/9303143}{{\ttfamily hep-th/9303143}}].

\bibitem{Shatashvili:1993ps}
S.~L. Shatashvili, \emph{{On the problems with background independence in
  string theory}}, \href{https://doi.org/10.1007/3-540-58453-6_12}{\emph{Alg.
  Anal.} {\bfseries 6} (1994) 215}
  [\href{https://arxiv.org/abs/hep-th/9311177}{{\ttfamily hep-th/9311177}}].

\bibitem{Friedan:2003yc}
D.~Friedan and A.~Konechny, \emph{{On the boundary entropy of one-dimensional
  quantum systems at low temperature}},
  \href{https://doi.org/10.1103/PhysRevLett.93.030402}{\emph{Phys. Rev. Lett.}
  {\bfseries 93} (2004) 030402}
  [\href{https://arxiv.org/abs/hep-th/0312197}{{\ttfamily hep-th/0312197}}].

\bibitem{Casini:2016fgb}
H.~Casini, I.~Salazar~Landea and G.~Torroba, \emph{{The g-theorem and quantum
  information theory}},
  \href{https://doi.org/10.1007/JHEP10(2016)140}{\emph{JHEP} {\bfseries 10}
  (2016) 140} [\href{https://arxiv.org/abs/1607.00390}{{\ttfamily
  1607.00390}}].

\bibitem{Rudelius:2020orz}
T.~Rudelius and S.-H. Shao, \emph{{Topological Operators and Completeness of
  Spectrum in Discrete Gauge Theories}},
  \href{https://doi.org/10.1007/JHEP12(2020)172}{\emph{JHEP} {\bfseries 12}
  (2020) 172} [\href{https://arxiv.org/abs/2006.10052}{{\ttfamily
  2006.10052}}].

\bibitem{Lauria:2020emq}
E.~Lauria, P.~Liendo, B.~C. Van~Rees and X.~Zhao, \emph{{Line and surface
  defects for the free scalar field}},
  \href{https://doi.org/10.1007/JHEP01(2021)060}{\emph{JHEP} {\bfseries 01}
  (2021) 060} [\href{https://arxiv.org/abs/2005.02413}{{\ttfamily
  2005.02413}}].

\bibitem{Collier:2021ngi}
S.~Collier, D.~Mazac and Y.~Wang, \emph{{Bootstrapping Boundaries and Branes}},
   \href{https://arxiv.org/abs/2112.00750}{{\ttfamily 2112.00750}}.

\bibitem{Herzog:2022jqv}
C.~P. Herzog and A.~Shrestha, \emph{{Conformal surface defects in Maxwell
  theory are trivial}},
  \href{https://doi.org/10.1007/JHEP08(2022)282}{\emph{JHEP} {\bfseries 08}
  (2022) 282} [\href{https://arxiv.org/abs/2202.09180}{{\ttfamily
  2202.09180}}].

\bibitem{Gimenez-Grau:2022czc}
A.~Gimenez-Grau, E.~Lauria, P.~Liendo and P.~van Vliet, \emph{{Bootstrapping
  line defects with O(2) global symmetry}},
  \href{https://doi.org/10.1007/JHEP11(2022)018}{\emph{JHEP} {\bfseries 11}
  (2022) 018} [\href{https://arxiv.org/abs/2208.11715}{{\ttfamily
  2208.11715}}].

\bibitem{Wilson:1974sk}
K.~G. Wilson, \emph{{Confinement of Quarks}},
  \href{https://doi.org/10.1103/PhysRevD.10.2445}{\emph{Phys. Rev. D}
  {\bfseries 10} (1974) 2445}.

\bibitem{Kaplan:2009kr}
D.~B. Kaplan, J.-W. Lee, D.~T. Son and M.~A. Stephanov, \emph{{Conformality
  Lost}}, \href{https://doi.org/10.1103/PhysRevD.80.125005}{\emph{Phys. Rev. D}
  {\bfseries 80} (2009) 125005}
  [\href{https://arxiv.org/abs/0905.4752}{{\ttfamily 0905.4752}}].

\bibitem{Gorbenko:2018ncu}
V.~Gorbenko, S.~Rychkov and B.~Zan, \emph{{Walking, Weak first-order
  transitions, and Complex CFTs}},
  \href{https://doi.org/10.1007/JHEP10(2018)108}{\emph{JHEP} {\bfseries 10}
  (2018) 108} [\href{https://arxiv.org/abs/1807.11512}{{\ttfamily
  1807.11512}}].

\bibitem{Benini:2019dfy}
F.~Benini, C.~Iossa and M.~Serone, \emph{{Conformality Loss, Walking, and 4D
  Complex Conformal Field Theories at Weak Coupling}},
  \href{https://doi.org/10.1103/PhysRevLett.124.051602}{\emph{Phys. Rev. Lett.}
  {\bfseries 124} (2020) 051602}
  [\href{https://arxiv.org/abs/1908.04325}{{\ttfamily 1908.04325}}].

\bibitem{pomeranchuk1945energy}
I.~Pomeranchuk and Y.~Smorodinsky, \emph{On the energy levels of systems with
  $z>137$}, {\emph{J. Phys. Ussr} {\bfseries 9} (1945) 97}.

\bibitem{Pereira_2007}
V.~M. {Pereira}, J.~{Nilsson} and A.~H. {Castro Neto}, \emph{{Coulomb Impurity
  Problem in Graphene}},
  \href{https://doi.org/10.1103/PhysRevLett.99.166802}{\emph{\prl} {\bfseries
  99} (2007) 166802} [\href{https://arxiv.org/abs/0706.2872}{{\ttfamily
  0706.2872}}].

\bibitem{shytov2007vacuum}
A.~V. {Shytov}, M.~I. {Katsnelson} and L.~S. {Levitov}, \emph{{Vacuum
  Polarization and Screening of Supercritical Impurities in Graphene}},
  \href{https://doi.org/10.1103/PhysRevLett.99.236801}{\emph{\prl} {\bfseries
  99} (2007) 236801} [\href{https://arxiv.org/abs/0705.4663}{{\ttfamily
  0705.4663}}].

\bibitem{Aharony:ToAppear}
O.~Aharony, G.~Cuomo, Z.~Komargodski, M.~Mezei and A.~Raviv-Moshe, \emph{{The
  Taxonomy of Wilson Lines}}, {\emph{To appear.} }.

\bibitem{Rodriguez-Gomez:2022gbz}
D.~Rodriguez-Gomez, \emph{{A scaling limit for line and surface defects}},
  \href{https://doi.org/10.1007/JHEP06(2022)071}{\emph{JHEP} {\bfseries 06}
  (2022) 071} [\href{https://arxiv.org/abs/2202.03471}{{\ttfamily
  2202.03471}}].

\bibitem{Aharony:2015afa}
O.~Aharony, G.~Gur-Ari and N.~Klinghoffer, \emph{{The Holographic Dictionary
  for Beta Functions of Multi-trace Coupling Constants}},
  \href{https://doi.org/10.1007/JHEP05(2015)031}{\emph{JHEP} {\bfseries 05}
  (2015) 031} [\href{https://arxiv.org/abs/1501.06664}{{\ttfamily
  1501.06664}}].

\bibitem{Witten:2001ua}
E.~Witten, \emph{{Multitrace operators, boundary conditions, and AdS / CFT
  correspondence}},  \href{https://arxiv.org/abs/hep-th/0112258}{{\ttfamily
  hep-th/0112258}}.

\bibitem{Klebanov:1999tb}
I.~R. Klebanov and E.~Witten, \emph{{AdS / CFT correspondence and symmetry
  breaking}}, \href{https://doi.org/10.1016/S0550-3213(99)00387-9}{\emph{Nucl.
  Phys. B} {\bfseries 556} (1999) 89}
  [\href{https://arxiv.org/abs/hep-th/9905104}{{\ttfamily hep-th/9905104}}].

\bibitem{Gubser:2002vv}
S.~S. Gubser and I.~R. Klebanov, \emph{{A Universal result on central charges
  in the presence of double trace deformations}},
  \href{https://doi.org/10.1016/S0550-3213(03)00056-7}{\emph{Nucl. Phys. B}
  {\bfseries 656} (2003) 23}
  [\href{https://arxiv.org/abs/hep-th/0212138}{{\ttfamily hep-th/0212138}}].

\bibitem{Faulkner:2009wj}
T.~Faulkner, H.~Liu, J.~McGreevy and D.~Vegh, \emph{{Emergent quantum
  criticality, Fermi surfaces, and AdS(2)}},
  \href{https://doi.org/10.1103/PhysRevD.83.125002}{\emph{Phys. Rev. D}
  {\bfseries 83} (2011) 125002}
  [\href{https://arxiv.org/abs/0907.2694}{{\ttfamily 0907.2694}}].

\bibitem{Diaz:2007an}
D.~E. Diaz and H.~Dorn, \emph{{Partition functions and double-trace
  deformations in AdS/CFT}},
  \href{https://doi.org/10.1088/1126-6708/2007/05/046}{\emph{JHEP} {\bfseries
  05} (2007) 046} [\href{https://arxiv.org/abs/hep-th/0702163}{{\ttfamily
  hep-th/0702163}}].

\bibitem{Giombi:2014xxa}
S.~Giombi and I.~R. Klebanov, \emph{{Interpolating between $a$ and $F$}},
  \href{https://doi.org/10.1007/JHEP03(2015)117}{\emph{JHEP} {\bfseries 03}
  (2015) 117} [\href{https://arxiv.org/abs/1409.1937}{{\ttfamily 1409.1937}}].

\bibitem{Gabai:2022vri}
B.~Gabai, A.~Sever and D.-l. Zhong, \emph{{Line Operators in
  Chern-Simons\textendash{}Matter Theories and Bosonization in Three
  Dimensions}},
  \href{https://doi.org/10.1103/PhysRevLett.129.121604}{\emph{Phys. Rev. Lett.}
  {\bfseries 129} (2022) 121604}
  [\href{https://arxiv.org/abs/2204.05262}{{\ttfamily 2204.05262}}].

\bibitem{Gabai:2022sij}
B.~Gabai, A.~Sever and D.-l. Zhong, \emph{{Line operators in
  Chern-Simons-Matter theories and Bosonization in Three Dimensions II
  -Perturbative Analysis and All-loop Resummation}},
  \href{https://arxiv.org/abs/2212.02518}{{\ttfamily 2212.02518}}.

\bibitem{Greiner:1985ce}
W.~Greiner, B.~Muller and J.~Rafelski, \emph{{Quantum Electrodynamics Of Strong
  Fields}}. 1985.

\bibitem{GARTNER1980181}
P.~G{\"a}rtner, B.~M{\"u}ller, J.~Reinhardt and W.~Greiner, \emph{Limiting
  charge for point nuclei},
  \href{https://doi.org/https://doi.org/10.1016/0370-2693(80)90464-5}{\emph{Physics
  Letters B} {\bfseries 95} (1980) 181}.

\bibitem{Polchinski:2011im}
J.~Polchinski and J.~Sully, \emph{{Wilson Loop Renormalization Group Flows}},
  \href{https://doi.org/10.1007/JHEP10(2011)059}{\emph{JHEP} {\bfseries 10}
  (2011) 059} [\href{https://arxiv.org/abs/1104.5077}{{\ttfamily 1104.5077}}].

\bibitem{Beccaria:2017rbe}
M.~Beccaria, S.~Giombi and A.~Tseytlin, \emph{{Non-supersymmetric Wilson loop
  in $ \mathcal{N} $ = 4 SYM and defect 1d CFT}},
  \href{https://doi.org/10.1007/JHEP03(2018)131}{\emph{JHEP} {\bfseries 03}
  (2018) 131} [\href{https://arxiv.org/abs/1712.06874}{{\ttfamily
  1712.06874}}].

\bibitem{Shnir:2005vvi}
Y.~M. Shnir, \emph{{Magnetic Monopoles}}, Text and Monographs in Physics.
  Springer, Berlin/Heidelberg, 2005,
  \href{https://doi.org/10.1007/3-540-29082-6}{10.1007/3-540-29082-6}.

\bibitem{Aharony:2013hda}
O.~Aharony, N.~Seiberg and Y.~Tachikawa, \emph{{Reading between the lines of
  four-dimensional gauge theories}},
  \href{https://doi.org/10.1007/JHEP08(2013)115}{\emph{JHEP} {\bfseries 08}
  (2013) 115} [\href{https://arxiv.org/abs/1305.0318}{{\ttfamily 1305.0318}}].

\bibitem{Kapustin:2005py}
A.~Kapustin, \emph{{Wilson-'t Hooft operators in four-dimensional gauge
  theories and S-duality}},
  \href{https://doi.org/10.1103/PhysRevD.74.025005}{\emph{Phys. Rev. D}
  {\bfseries 74} (2006) 025005}
  [\href{https://arxiv.org/abs/hep-th/0501015}{{\ttfamily hep-th/0501015}}].

\bibitem{Pisarski:1984dj}
R.~D. Pisarski, \emph{{Chiral Symmetry Breaking in Three-Dimensional
  Electrodynamics}},
  \href{https://doi.org/10.1103/PhysRevD.29.2423}{\emph{Phys. Rev. D}
  {\bfseries 29} (1984) 2423}.

\bibitem{kotov2012electron}
V.~N. {Kotov}, B.~{Uchoa}, V.~M. {Pereira}, F.~{Guinea} and A.~H. {Castro
  Neto}, \emph{{Electron-Electron Interactions in Graphene: Current Status and
  Perspectives}},
  \href{https://doi.org/10.1103/RevModPhys.84.1067}{\emph{Reviews of Modern
  Physics} {\bfseries 84} (2012) 1067}
  [\href{https://arxiv.org/abs/1012.3484}{{\ttfamily 1012.3484}}].

\bibitem{wang2013observing}
Y.~{Wang}, D.~{Wong}, A.~V. {Shytov}, V.~W. {Brar}, S.~{Choi}, Q.~{Wu} et~al.,
  \emph{{Observing Atomic Collapse Resonances in Artificial Nuclei on
  Graphene}}, \href{https://doi.org/10.1126/science.1234320}{\emph{Science}
  {\bfseries 340} (2013) 734}
  [\href{https://arxiv.org/abs/1510.02890}{{\ttfamily 1510.02890}}].

\end{thebibliography}\endgroup
\bibliographystyle{JHEP.bst}

\end{document}